\documentclass[prl,aps,showpacs,twocolumn,floatfix,superscriptaddress,showkeys]{revtex4}  
\usepackage{graphicx}
\usepackage[ansinew]{inputenc}
\usepackage{array}
\usepackage{color}
\usepackage{amsmath}
\usepackage{amsxtra}
\usepackage{amstext}
\usepackage{amssymb}
\usepackage{latexsym}
\usepackage{dsfont}
\usepackage{verbatim}
\usepackage{comment}
\usepackage{epstopdf}

\begin{document}

\title{A method for achieving larger enhancement in Four-Wave Mixing \\ via plasmonic path interference effects}

\author{Shailendra K. Singh}
\affiliation{Institute of Nuclear Sciences, Hacettepe University, 06800, Ankara, Turkey}
\author{M. Kurtulus Abak  }
\affiliation{ GUNAM, Ortado\u{g}u Teknik \"{U}niversitesi,  06800, Ankara, Turkey}
\author{Mehmet Emre Tasgin }
\affiliation{Institute of Nuclear Sciences, Hacettepe University, 06800, Ankara, Turkey}
\affiliation{to whom correspondence should be addressed: metasgin@hacettepe.edu.tr}

\date{\today}

\begin{abstract}
Enhancement and suppression of nonlinear processes in coupled systems of plasmonic converters and quantum emitters are well-studied theoretically, numerically and experimentally, in the past decade. Here, in difference, we explicitly demonstrate --with a single equation-- how the presence of a Fano resonance leads to cancellation of nonresonant terms in a four-wave mixing process. Cancellation in the denominator gives rise to enhancement  in the nonlinearity. The explicit demonstration, we present here, guides us to the method for achieving more and more orders of magnitude enhancement factors via path interference effects. We also study the coupled system of a plasmonic converter with two quantum emitters. We show that the potential for the enhancement increases dramatically due to better cancellation of the terms in the denominator.
\end{abstract}

\pacs{42.50.Gy, 78.67.Bf, 42.65.Hw, 73.20.Mf}

\keywords{Four-wave mixing,  Fano resonances, plasmons, enhancement}



\maketitle


{\it Introduction.} Plasmonic resonators, such as metal nanoparticles (MNPs), graphene nano-islands \cite{Abajo-ACS-2015-graphene} and  transparent conducting oxides \cite{TCO}, confine the incident optical electromagnetic field into nano-dimensions in the form of surface plasmon (SP) excitations. When a quantum emitter (QE) is placed on one of these hot-spots, plasmons interact with the emitter strongly \cite{QDattached}. Due to the small decay rate of the QE, Fano resonances appear in the spectrum of the plasmonic material, where absorption vanishes. Fano resonances, appearing in the linear response \cite{QCfano1}, can extend the lifetime of plasmons \cite{metasginNanoscale2013} and lead to further enhancement of the local field \cite{FanoRes-NatureMat-2010} which makes coherent plasmon emission (spaser) possible \cite{spaser}. 

Linear Fano resonances do not emerge only when a plasmon is coupled with a QE, but also observed for two interacting plasmonic materials if one has a smaller decay rate \cite{Soukoulis2012}. The origin of Fano resonance is analogical to electromagnetically induced transparency (EIT) \cite{Scullybook}, and can clearly be understood in terms of interference of two absorption paths \cite{Alzar}.

In the past decade, the role of Fano resonances in the modification of nonlinear response is also well-studied. Presence of a  quantum emitter, with an energy spacing $\omega_{eg}$ close to the output frequency, can enhance or silence the frequency conversion \cite{Walsh&Negro-NanoLett-2013,silencing_enhancementSHG,SHG-MNP+molecules-2000,MartinNanoLett2013,BrevetnonlinearFanoPRB2012,Martin-OptExp-2014,Tasgin-JOpt-2014}. Both effects, suppression and enhancement, are desirable for devices designed to operate in the linear and nonlinear regimes, respectively. Operation of high-power lasers and fiber optic cables \cite{avoidNL1} necessitates the avoidance of second harmonic generation (SHG) and Raman scattering, in order to prevent the loss of energy to other modes. On the other hand, nonlinear imaging with SHG \cite{SHGimaging}, four-wave mixing (FWM) \cite{Bartal-PRL-2014-FWMresolution} and Raman spectroscopy \cite{Sharma2012}, and generation of nonclassical/entangled plasmons \cite{plasmon-entanglement} desire enhanced nonlinear response.

Four-wave mixing (FWM) process --beyond its technical applications such as superresolution imaging \cite{Bartal-PRL-2014-FWMresolution}, ultrafast optical switching \cite{plasmon-switch-FWM} and nonlinear negative refraction \cite{Zhang-NatureMat-2012}-- is also important for studying one of the fundamental inquiries in quantum information. In a FWM process, two fields become entangled in addition to squeezing in one of the fields. Hence, FWM is a special process in which the interplay between single-mode nonclassicality and two-mode entanglement coexist naturally \cite{nonclassicality}. Localization is shown to enhance also FWM in plasmonic materials \cite{Novotny-PRL-2009} due to the larger overlap in mode-integrals for that process \cite{ZayatsPRB2012,overlapintegral}.

Similar to SHG \cite{SHG-MNP+molecules-2000}, presence of quantum emitters attached to plasmonic converters modifies the FWM process \cite{Terzis-JAppPhys-2014,Zhu-JPhysB-2008,Paspalakis-2014,Wang-OptExpress-2012,Halas-PNAS-2013}. Analogous to linear response \cite{Soukoulis2012} and SHG \cite{Walsh&Negro-NanoLett-2013,silencing_enhancementSHG,BrevetnonlinearFanoPRB2012,Martin-OptExp-2014}, coupling of two plasmonic materials with different decay rates creates modifications due to Fano resonances also in the FWM process \cite{classical_FWM_enhance}. Fano resonances in FWM can show themselves even as the coupling between superradiant and subradiant collective modes of plasmonic nanoclusters \cite{Halas-PNAS-2013}, similar to dark states in a plasmonic SHG process \cite{Remo_2014}.

Presence and behavior of such resonances are well-studied in the literature. In this Letter, in difference, we demonstrate the path interference effects {\it explicitly} in a plasmonic FWM process. We provide a single equation [see Eq.~(\ref{alp3})] for the steady state amplitude of the FWM conversion. Interaction with the quantum emitter introduces extra terms in the denominator of the conversion amplitude. We show that enhancement emerges simply due to the cancellation of these extra terms with the nonresonant term in the denominator. On the contrary, conversion suppression emerges when the extra term grows several orders of magnitude --due to the small quantum decay rate-- and makes the denominator blow up. 

We also study the system where a plasmonic converter is coupled to two quantum emitters (QEs). We show that a better cancellation of the nonresonant term --this time also the decay term-- can be achieved in the FWM process, see Eq.~(\ref{alp3_2QE}). FWM enhancement increases by an order of magnitude (15 times) compared to the system where the plasmonic converter is coupled with a single QE.

We present the second quantized Hamiltonian for the system of a plasmonic converter (FWM) coupled to a single quantum emitter. We obtain the equations of motion, introduce decay terms, convert operators to c-numbers (since not interested in entanglement) and obtain the steady-state result for the converted FWM frequency [see Eq. (\ref{alp3})]. We repeat the similar procedure for the system of plasmonic converter coupled to two QEs. In a previous publication we showed that one does not need to concern about the retardation effects. We simulated SHG process in a coupled system using 3D Maxwell equations and we observed that the spectral position and the strength of Fano enhancement are not affected significantly (see Fig.~5 in Ref. \cite{Tasgin-JOpt-2014}).

\begin{figure}[t]
\centering
\includegraphics[width=0.4\textwidth]{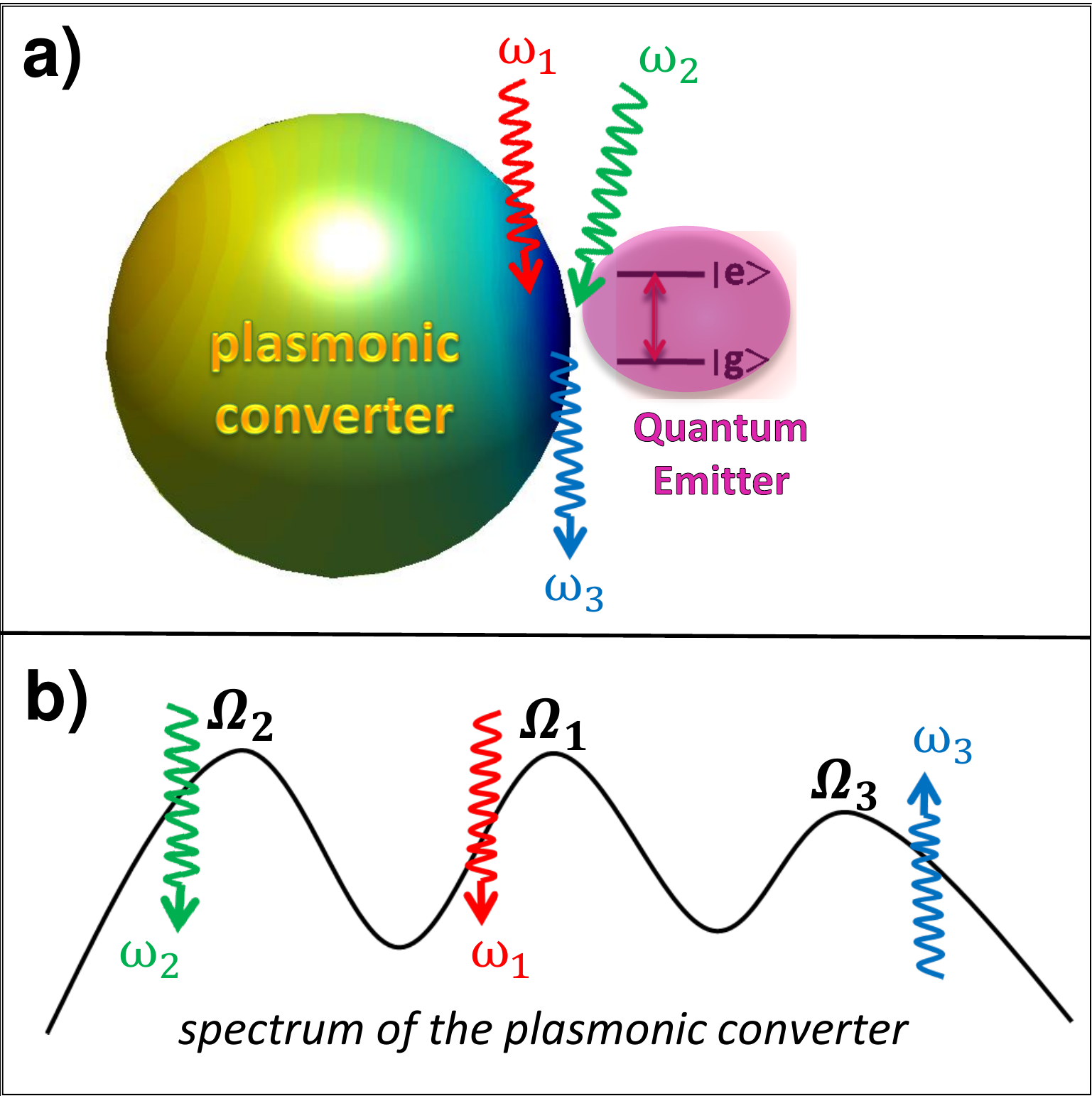}
\caption{\label{fig1} (Color online) (a) A quantum emitter (QE) is placed to the hot-spot of a plasmonic converter. Field localization provides a strong plasmon-QE interaction. (b) Two lasers of frequencies $\omega_1$ and $\omega_2$ drive the plasmon modes of resonances $\Omega_1$ and $\Omega_2$, respectively. 2 plasmons in the $\Omega_1$ mode, oscillating with $\omega_1$, combine and decay into two plasmons with different frequencies, one into $\Omega_2$ mode and another one into the $\Omega_3$ mode. Decay into the $\Omega_2$ mode is stimulatd by the $\omega_2$ laser. Frequencies are scaled with $\omega_1$.}
\label{fig1}
\end{figure}

{\it Hamiltonian}. Nonlinear processes in plasmonic converters take place through plasmons \cite{2plas1phot} due to growing mode (overlap) integrals \cite{ZayatsPRB2012,overlapintegral} with the localization. In FWM process, $\hat{a}_3^\dagger\hat{a}_2^\dagger\hat{a}_1^2 + H.c.$, the mode integral determing the strength of this process is proportional as \cite{ZayatsPRB2012}
\begin{equation}
\chi_{\scriptscriptstyle{\rm FWM}}\sim  \int d^3{\bf r} \: E_3^*({\bf r}) E_2^*({\bf r}) E_1^2({\bf r}) \; ,
\label{chi}
\end{equation}
where $E_i$ are the spatial extents of the electric (or polarization) fields of the $\hat{a}_i$ plasmon modes around the nanoparticle. A nonvanishing $\chi_{\scriptscriptstyle{\rm FWM}}$ necessitates some parity requirements for the fields of the plasmon modes, in symmetric particles.

The dynamics of the conversion process can be describes as follows. Plasmonic converter has three resonances, of frequencies $\Omega_1$, $\Omega_2$, and $\Omega_3$, in the relevant region of the spectrum (see Fig.~\ref{fig1}b). Two lasers of frequencies $\omega_1$ and $\omega_2$ drive the two plasmon modes  $\Omega_1$, $\Omega_2$, respectively. The lasers excite surface plasmons in the $\hat{a}_1$-mode ($\Omega_1$) and $\hat{a}_2$-mode ($\Omega_2$), wherein plasmons oscillate at $\omega_1$ and $\omega_2$. 2 plasmons in the $\Omega_1$ mode combine and decay into two plasmons with different frequencies, one in the $\Omega_2$ mode and another one in the $\Omega_3$ mode. The plasmon generated in the $\hat{a}_3$ mode ($\Omega_3$) oscillates with frequency $\omega_3=2\omega_1-\omega_2$. Decay into $\Omega_2$ mode is stimulated by the presence of plasmons in this mode. $\omega_3$ plasmon oscillations emerge due to the energy conservation and the presence of a plasmon mode at $\Omega_3 \sim \omega_3$. In order to observe the $\omega_3$ photons in the far field, plasmons in the $\Omega_3$ mode can be transformed to photons selectively by gratings \cite{Novotny-PRL-2009}. 

Therefore, the Hamiltonian for the total system can be written as the sum of the energies of the quantum emitter ($\hat{H}_0$), plasmon modes ($\hat{H}_{\rm pls}$), the plasmon-QE interaction ($\hat{H}_{\rm int}$)
\begin{equation}
\hat{H}_{0}=\hbar \omega _{e}\left\vert e\right\rangle \left\langle
e\right\vert +\hbar \omega _{g}\left\vert g\right\rangle \left\langle
g\right\vert \; ,
\label{H0}
\end{equation}
\begin{equation}
\hat{H}_{\rm pls}=\hbar \Omega _{1}\hat{a}_{1}^{\dagger }\hat{a}_{1}+\hbar
\Omega _{2}\hat{a}_{2}^{\dagger }\hat{a}_{2}+\hbar \Omega _{3}\hat{a}%
_{3}^{\dagger }\hat{a}_{3} \; ,
\end{equation}
\begin{equation}
\hat{H}_{int}=\hbar \left( f\hat{a}_{3}^{\dagger }\left\vert g\right\rangle
\left\langle e\right\vert +f^*\hat{a}_{3}\left\vert e\right\rangle
\left\langle g\right\vert \right) \; ,
\end{equation}
and including the two laser pumps ($\hat{H}_{\rm pump}$) and the nonlinear FWM process ($\hat{H}_{\scriptscriptstyle{\rm FWM}}$)
\begin{equation}
\hat{H}_{p}=i\hbar \left( \hat{a}_{1}^{\dagger }~\varepsilon _{\rm p}^{(1)}e^{-i\omega_1 t} + 
\hat{a}_{2}^{\dagger }~\varepsilon _{\rm p}^{(2)}e^{-i\omega_2 t} \; + \; H.c. 
\right) \; ,
\end{equation}
\begin{equation}
\hat{H}_{\scriptscriptstyle{\rm FWM}}=\hbar \chi_{\scriptscriptstyle{\rm FWM}} \left( \hat{a}_{3}^{\dagger }\hat{a}%
_{2}^{\dagger }\hat{a}_{1}^{2}+\hat{a}_{1}^{\dagger 2}\hat{a}_{2}\hat{a}%
_{3}~\right) \; ,
\label{Hfwm}
\end{equation}
where $f$ is the strength for plasmon-QE interaction, $\omega_e$ ($\omega_g$) is the excited (ground) state of the QE, $\varepsilon_{\rm p}^{(1)}$ and $\varepsilon_{\rm p}^{(2)}$ are the strengths of the two lasers of frequency $\omega_1$ and $\omega_2$, and mode-integral of the nonlinear process $\chi_{\scriptscriptstyle{\rm FWM}}$ is defined in Eq.~(\ref{chi}). We choose the energy level spacing of the QE, $\omega_{eg}=\omega_e-\omega_g$, such that it falls into the spectral region of the $\Omega_3$ plasmon mode. Hence, we consider the interaction of QE with $\hat{a}_3$ mode only, which provides substantial simplification making analytic results [Eqs.~(\ref{alp3}) and (\ref{alp3_2QE})] possible \cite{PS1}.

{\it Equations of Motion.} Dynamics of the plasmon operators $\hat{a}_i$ and the density operators for the QE $\rho_{eg}=|e\rangle\langle g|$ and $\quad \rho_{ee}=|e\rangle\langle e|$ can be obtained from the Heisenberg equation of motion, e.g. $i\hbar\dot{\hat{a}}_i=[\hat{a}_i,\hat{H}]$. We are not interested in the correlations in the system. Oscillation amplitudes (or occupation numbers) of the plasmon modes can as well be handled by replacing operators by $c$-numbers, i.e. $\hat{a}_i \rightarrow \alpha_i$. When we introduce the plasmonic linewidths $\gamma_1$, $\gamma_2$, $\gamma_3$ and the decay rate of the quantum emitter $\gamma_{eg}$, dynamics is governed by coupled equations
\begin{subequations}
\begin{equation}
\dot{\alpha}_{1}=\left( -i\Omega_{1}-\gamma _{1}\right) \alpha _{1}-i2 \chi_{\scriptscriptstyle{\rm FWM}} 
\alpha _{1}^* \alpha _{2}\alpha _{3}+\varepsilon_{\rm p}^{(1)}e^{-i\omega_1 t}  \label{1QEa}
\end{equation}
\begin{equation}
\dot{\alpha}_{2}=\left( -i\Omega_{2}-\gamma _{2}\right) \alpha _{2}-i \chi_{\scriptscriptstyle{\rm FWM}} 
\alpha _{3}^* \alpha _{1}^{2}+\varepsilon _{\rm p}^{(2)} e^{-i\omega_2 t}  \label{7b}
\end{equation}
\begin{equation}
\dot{\alpha}_{3}=\left( -i\Omega _{3}-\gamma _{3}\right) \alpha _{3}-i\chi_{\scriptscriptstyle{\rm FWM}} 
\alpha _{2}^* \alpha _{1}^{2}-if\rho_{ge}
\label{7c}
\end{equation}
\begin{equation}
\dot{\rho}_{ge}=\left( -i\omega _{eg}-\gamma _{eg}\right) \rho
_{ge}+if\alpha _{3}\left( \rho _{ee}-\rho _{gg}\right)  \label{7d}
\end{equation}
\begin{equation}
\dot{\rho}_{ee}=-\gamma _{ee}\rho _{ee}+~i\left( f\alpha_{3}^*\rho_{ge}- 
f^*\alpha_{3}\rho _{eg}\right)  \label{1QEe}
\end{equation}
\end{subequations}
with $\rho_{gg}=1-\rho_{ee}$ and $\gamma_{ee}=2\gamma_{eg}$.

{\it Steady-state.} When the dynamics of Eqs.~(\ref{1QEa})-(\ref{1QEe}) is examined by placing exponential solutions, one identifies 
\begin{equation}
\alpha_i(t)=\tilde{\alpha}_i e^{-i\omega_i t}  \quad \text{and} \quad \rho_{ge}=\tilde{\rho}_{ge} e^{-i(2\omega_1-\omega_2)t}
\label{exponen}
\end{equation}
as the steady-state oscillations, where $\tilde{\alpha}_i$ and $\tilde{\rho}_{ge}$ are constants. It is worth noting that, in order to be confident about the validity of Eq.~(\ref{exponen}), we obtain Figs.~2 and 3 by time evolving Eqs.~(\ref{1QEa})-(\ref{1QEe}). When we place Eq.~(\ref{exponen}) into Eqs.~(\ref{1QEa})-(\ref{1QEe}), we reach the coupled nonlinear equations
\begin{subequations}
\begin{equation}
\left[ i\left( \Omega _{1}-\omega_1 \right) +\gamma _{1}\right] \tilde{\alpha}%
_{1}+i2\chi_{\scriptscriptstyle{\rm FWM}} \tilde{\alpha}_{1}^{\ast }\tilde{\alpha}_{2}%
\tilde{\alpha}_{3}=\varepsilon_{p}^{(1)}  \label{steady_a}
\end{equation}
\begin{equation}
\left[ i\left( \Omega_{2}-\omega_2 \right) +\gamma _{2}\right] 
\tilde{\alpha}_{2}+i\chi_{\scriptscriptstyle{\rm FWM}} \tilde{\alpha}_{3}^{\ast }\tilde{%
\alpha}_{1}^{2}=\varepsilon _{p}^{(2)}  \label{steady_b}
\end{equation}
\begin{equation}
\left[ i\left( \Omega _{3}+\omega_2-2\omega_1 \right) +\gamma _{3}%
\right] \tilde{\alpha}_{3}+i\chi_{\scriptscriptstyle{\rm FWM}} \tilde{\alpha}_{2}^{\ast
}\tilde{\alpha}_{1}^{2}=-if\tilde{\rho}_{ge}  \label{steady_c}
\end{equation}
\begin{equation}
\left[ i\left( \omega _{eg}+\omega_2-2\omega_1 \right) +\gamma _{eg}%
\right] \tilde{\rho}_{ge}=if\tilde{\alpha}_{3}\left( {\rho}_{ee}-%
{\rho}_{gg}\right)  \label{steady_d}
\end{equation}
\begin{equation}
\gamma _{ee}\tilde{\rho}_{ee}=i\left( f\tilde{\alpha}_{3}^{\ast }\tilde{\rho}%
_{ge} - f^*\tilde{\alpha}_{3}\tilde{\rho}_{eg}\right)  \label{steady_e}
\end{equation}
\end{subequations}
for the steady-state plasmon occupations, $|\tilde{\alpha}_i|^2$

{\it The single equation.} With a simple algebraic manipulation, using Eqs.~(\ref{steady_c}) and (\ref{steady_d}), one can reach the very simple and useful equation
\begin{equation}
\tilde{\alpha}_{3}=\frac{i\chi_{\scriptscriptstyle{\rm FWM}} \tilde{\alpha}_{2}^{\ast }%
\tilde{\alpha}_{1}^{2}}{\frac{\left\vert f\right\vert ^{2}y}{\left[ i\left(
\omega _{eg}+\omega_2-2\omega_1 \right) +\gamma _{eg}\right] }-\left[
i\left( \Omega_{3}+\omega_2-2\omega_1 \right) +\gamma _{3}\right] }
\label{alp3}
\end{equation}
which determines the number of FWM plasmons $|\tilde{\alpha}_3|^2$, with $y=\rho_{ee}-\rho_{gg}$ is the population inversion for the QE. When coupling between the  plasmonic converter and the QE is not present, $f=0$, Eq.~(\ref{alp3}) displays the simple resonance condition for FWM. Maximum conversion is attained when $2\omega_1-\omega_2=\Omega_3$, as should be expected. This happens when the converted frequency is resonant with the plasmon mode. In Eq.~(\ref{alp3}) we assume resonant pumping for the two lasers, $\omega_1=\Omega_1$ and $\omega_2=\Omega_2$, for simplicity. We refer the term $\Omega_3-(2\omega_1-\omega_2)$, present in the denominator of (\ref{alp3}), as the nonresonant term. FWM is a very weak process. Hence, one can safely consider that occupations of the two plasmon modes, $|\tilde{\alpha}_1|^2$ and $|\tilde{\alpha}_2|^2$, are almost unaffected from the conversion process. In most cases, population stays weakly excited and one can imagine it as $y\simeq -1$.

\begin{figure}
\centering
\includegraphics[width=0.52\textwidth]{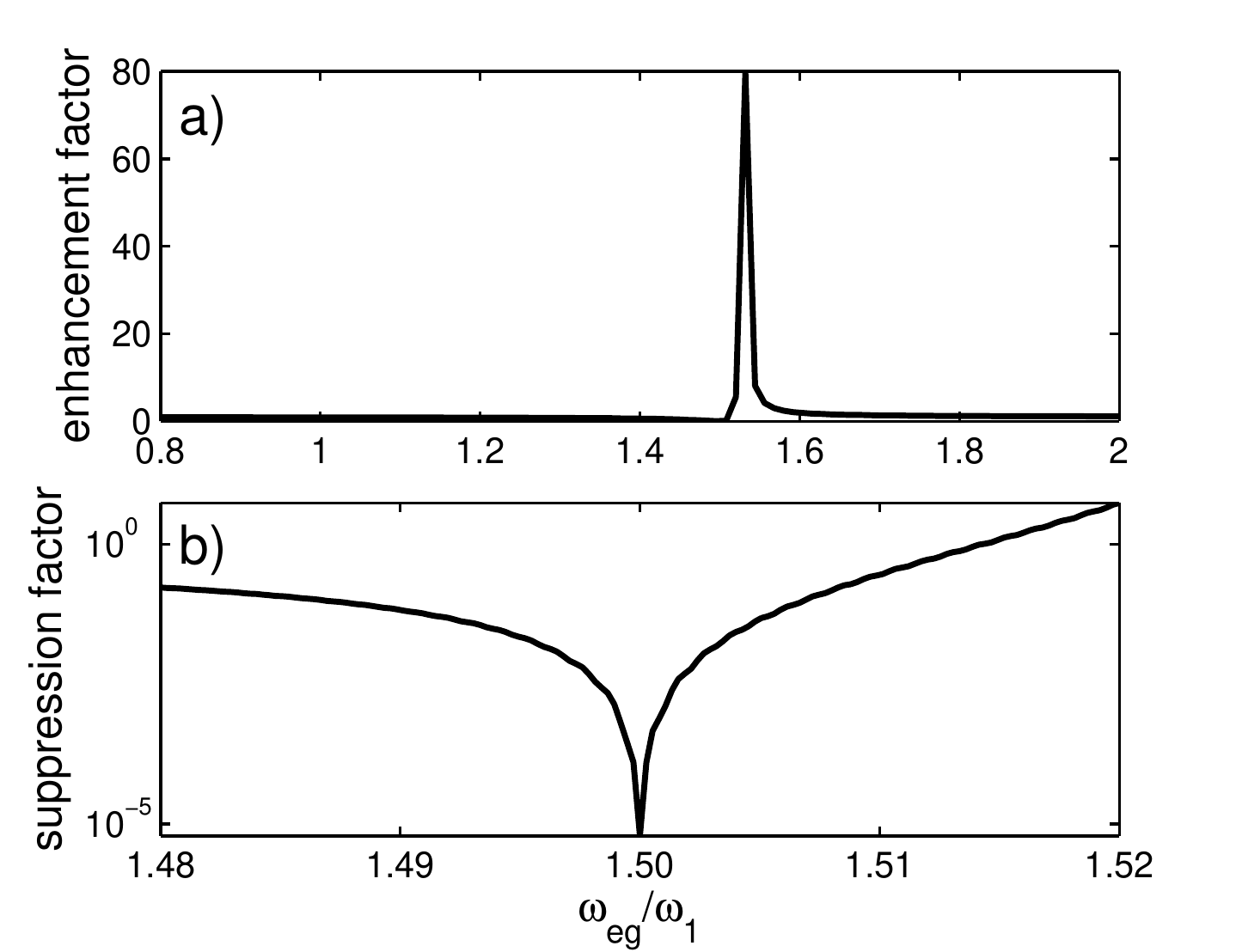}
\caption{\label{fig2} (a) Relative enhancement of the FWM process in the presence of coupling to the QE, compared to bare plasmonic converter. The firs term in the denominator of Eq.~(\ref{alp3}) cancels the following nonresonant term. Such a cancellation occurs for $\omega_{eg}\simeq 1.53$ [Eq.~\ref{weg}] which yields an 80 times enhancement. (b) When $\omega_{eg}=2\omega_1-\omega_2=\omega_3$, the extra term in the denominator becomes $y|f|^2/\gamma_{eg}$ very large and suppresses FWM.}
\label{fig2}
\end{figure}

{\it Enhancement} in FWM, in $|\tilde{\alpha}_3|^2$, can be achieved if one arranges the imaginary part of the first term of the denominator, $\left\vert f\right\vert ^{2}y /{\left[ i\left(
\omega _{eg}+\omega_2-2\omega_1 \right) +\gamma _{eg}\right] }$, to cancel the nonresonant term $(\Omega_{3}+\omega_2-2\omega_1)$. Therefore, if the level spacing of the QE is chosen such that
\begin{equation}
\omega_{eg}^*= \omega_3 + \frac{|f|^2y}{\Omega_3-\omega_3} + \sqrt{\frac{|f|^4y^2}{(\Omega_3-\omega_3)^2} - 4\gamma_{eg}^2}
\label{weg}
\end{equation}
there emerges an enhancement peak in the FWM spectrum, where $\omega_3=2\omega_1-\omega_2$. This is depicted in Fig.~\ref{fig2}a. Frequencies are scaled with $\omega_1$ and $\omega_1=\Omega_1=1$, $\omega_2=\Omega_2=0.5$, $\omega_3=1.5$, $\Omega_3=1.85$, $\chi_{\scriptscriptstyle{\rm FWM}}=10^{-5}$ $f=0.1$, $\gamma_1=\gamma_2=\gamma_3=0.01$, $\gamma_{eg}=10^{-5}$.

{\it Suppression} of FWM, on the contrary, can be obtained by arranging the term $\left\vert f\right\vert ^{2}y /{\left[ i\left(
\omega _{eg}+\omega_2-2\omega_1 \right) +\gamma _{eg}\right] }$ to blow up. This happens when the level spacing of the QE is set to $\omega_{eg}=2\omega_1-\omega_2$. Since $\gamma_{eg}\sim 10^9$Hz and plasmon resonances are typically $\sim 10^{15}$Hz, the extra term becomes very very large and leads to vanishing FWM, $|\tilde{\alpha}_3|^2$. This is depicted in Fig.~\ref{fig2}b.

{\it More and more enhancement.} The enhancement in Eq.~(\ref{alp3}) is limited by the presence of $\gamma_3$, which is not a small quantity. Hence, one naturally seeks a way to introduce more extra terms to get rid of (cancel) the decay term $\gamma_3$ in the denominator. For this purpose, we consider the system where the plasmonic converter is coupled to two QEs, with strengths  $f_1$ and $f_2$. The two QEs also interact with each other with a strength $g$.

Hamiltonian (\ref{H0})-(\ref{Hfwm}) needs to be modified with additional and replacing terms
\begin{equation}
\hat{H}_{\scriptscriptstyle{\rm QE-QE}}=\hbar \left( g \left\vert e_{2}\right\rangle
\left\langle g_{2}\right\vert \otimes \left\vert g_{1}\right\rangle
\left\langle e_{1}\right\vert  \; + \; H.c. \right) \; ,
\end{equation}
\begin{equation}
\hat{H}_{\rm int}=\hbar \left( f_{1}\hat{a}_{3}^{\dagger }\left\vert
g_{1}\right\rangle \left\langle e_{1}\right\vert  + 
f_{2}\hat{a}_{3}^{\dagger }\left\vert
g_{2}\right\rangle \left\langle e_{2}\right\vert 
\; + \; H.c. \right) \; .
\end{equation}
When the procedure, similar to Eqs.~(\ref{1QEa})-(\ref{1QEe}) and (\ref{steady_a})-(\ref{steady_e}), is applied \cite{supplementary} one again reach a single equation
\begin{widetext}
\begin{equation}
\tilde{\alpha}_{3}=\frac{i\chi_{\scriptscriptstyle{\rm FWM}}\left( \beta _{1}\beta
_{2}+y_{1}y_{2}\left\vert g\right\vert ^{2}\right) }{\left( y_{1}\left\vert
f_{1}\right\vert ^{2}\beta _{2}+y_{2}\left\vert f_{2}\right\vert ^{2}\beta
_{1}\right) +iy_{1}y_{2}\left( f_{1}f_{2}^{\ast }g^{\ast }+f_{1}^{\ast
}f_{2}g\right) -\xi_{3}\left( \beta _{1}\beta
_{2}+y_{1}y_{2}\left\vert g\right\vert ^{2}\right) }\tilde{\alpha}_{2}^{\ast
}\tilde{\alpha}_{1}^{2} \; ,
\label{alp3_2QE}
\end{equation}
\end{widetext}
where $\xi_{3}=i\left(\Omega_{3}+\omega_2-2\omega_1 \right) +\gamma_{3}$\ and  $\beta_{j}= i\left(\omega _{eg}^{( j )}+\omega_2-2\omega_1 \right) +\gamma _{eg}^{( j)}$ with $j=1,2$ enumerates QEs. Here, $\omega_{eg}^{(j)}$ and $\gamma_{eg}^{(j)}$ are the energy level spacing and the decay rate for the $j^{\rm th}$ QE. $y_1$ and $y_2$ are population inversions. 

This time, the extra terms cancel $\gamma_3$ more efficiently. In Fig.~\ref{fig3}, we see that the cancellation in the denominator [Eq.~(\ref{alp3_2QE})] results 1200 time enhancement for FWM, which is 15 times larger compared to the one for single QE [Eq.~\ref{alp3}]. In order to maximize $\tilde{\alpha}_3$ in Eq.~(\ref{alp3_2QE}), we numerically minimize the denominator by varying $f_1$, $f_2$, $g$, $\omega_{eg}^{(1)}$ and $\omega_{eg}^{(2)}$ \cite{supplementary}. In principle, the desired phase for the QE-QE interaction ($g$) or plasmon-QE interactions ($f_{1,2}$) can be obtained by arranging the phase of the interaction integral \cite{overlapintegral} by reshaping the two QEs. In the case of all-plasmonic Fano resonances \cite{silencing_enhancementSHG,MartinNanoLett2013} such an arrangement among the interacting plasmon modes would be much simpler. 

\begin{figure}
\centering
\includegraphics[width=0.49\textwidth]{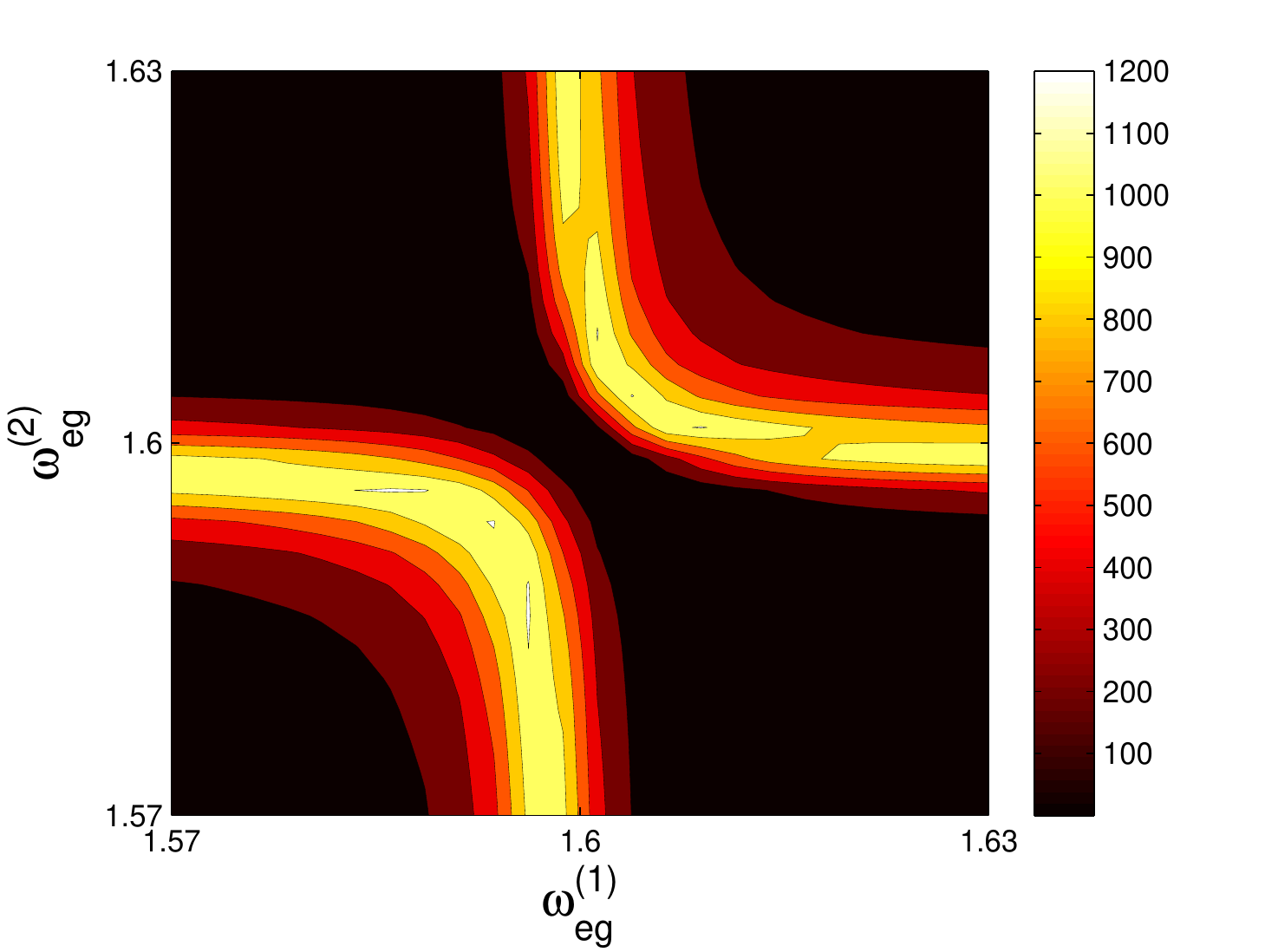}
\caption{\label{fig3} (Color online) When the plasmonic converter is coupled to 2 QEs, there introduces additional terms, see Eq.~(\ref{alp3_2QE}). These terms manage to partially cancel also the $\gamma_3$ term, which is not possible using a single QE as in Eq.~(\ref{alp3}). Enhancement reaches to 1200 times. }
\label{fig3}
\end{figure}

{\it Conclusion.} We obtain the steady-state amplitude of FWM process as a single equation for a plasmonic converter coupled to quantum emitters (QEs). The denominator of this equation reveals the path interference effects explicitly. Presence of QEs, interacting with the plasmonic converter, introduces extra terms in the denominator. When the extra terms cancel nonresonant term for the conversion, enhancement of FWM is observed. Contrary, when the extra term blows up due to the long lifetime of the QE, suppression of FWM is observed.

We utilize this observation as a tool for obtaining larger and larger enhancement in FWM. We show that, for the coupling of the plasmonic converter to two QEs, a better cancellation of the denominator is achieved. Enhancement increases by a factor of 15 on top of the resonance obtained by coupling with a single QE. Fano resonances, both in the linear and nonlinear response, emerge also by coupling the plasmonic converter with a plasmonic material which has a smaller decay rate \cite{silencing_enhancementSHG,MartinNanoLett2013,BrevetnonlinearFanoPRB2012,Martin-OptExp-2014}. Hence, our method is possible to be generalized \cite{classicalSHGenhance} to interacting plasmonic clusters, since the experiments are easier to conduct with nanoparticles.

\begin{acknowledgements}
M.E.T and S.K.S. acknowledge  support  from T\"{U}B\.{I}TAK-1001  Grant No. 114F170.
\end{acknowledgements}


\end{document}